\documentclass[showpacs,prl,superscriptaddress,twocolumn]{revtex4-1}
\usepackage{graphicx}
\usepackage{enumerate}
\usepackage{hyperref}
\usepackage{textcomp}
\usepackage{amsmath}
\usepackage{amssymb}
\usepackage{pgf}
\usepackage{relsize}
\usepackage{pgfpages}
\usepackage{pgfplots}
\usepackage{tikz}
\usepackage{xifthen}
\usetikzlibrary{quotes,angles}
\usetikzlibrary{arrows}
\usetikzlibrary{shapes.misc,shadows}
\usetikzlibrary{shapes,arrows}
\usetikzlibrary{calc,decorations.markings}
\usetikzlibrary{patterns}
\definecolor{myred}{rgb}{0.7,0.1,0.1}
\definecolor{myblue}{rgb}{0,0.447,0.741}
\definecolor{mygreen}{rgb}{0,0.498,0}
\definecolor{silvergray}{rgb}{0.752941176,0.752941176,0.752941176}
\definecolor{CateRed}{RGB}{161,24,1}

\graphicspath{{Figures/PNG/}{Figures/PDF/}{Figures/}}

\def\bra#1{\langle#1 |}
\def\ket#1{| #1\rangle}

\newcommand{\nep}{\textrm{e}}

\newcommand{\LE}{\mathrm{\scriptscriptstyle LE}}

\newcommand{\Tr}{{\rm Tr}}

\newcommand{\opadag}[1]{{\hat{a}^{\dagger}}_{#1}}
\newcommand{\opa}[1]{{\hat{a}^{\phantom \dagger}}_{#1}}
\newcommand{\opbdag}[1]{{\hat{b}^{\dagger}}_{#1}}
\newcommand{\opb}[1]{{\hat{b}^{\phantom \dagger}}_{#1}}
\newcommand{\opfdag}[1]{{\hat{f}^{\dagger}}_{#1}}
\newcommand{\opf}[1]{{\hat{f}^{\phantom \dagger}}_{#1}}

\newcommand{\Ham}{\hat{H}}

\begin{document}
	
	\title{Non-adiabatic breaking of topological pumping}

\author{Lorenzo Privitera}
\affiliation{SISSA, Via Bonomea 265, I-34136 Trieste, Italy}
\author{Angelo Russomanno}
\affiliation{Scuola Normale Superiore, Piazza dei Cavalieri 7, I-56127 Pisa, Italy}
\affiliation{International Centre for Theoretical Physics (ICTP), P.O.Box 586, I-34014 Trieste, Italy}

\author{Roberta Citro}
 \affiliation{Dipartimento di Fisica ``E.R. Caianiello'', Universit\`a
  degli Studi di Salerno and Spin-CNR Unit\`a, Via Giovanni Paolo II, 132, I-84084 Fisciano
  (Sa), Italy}

\author{Giuseppe E. Santoro}
\affiliation{SISSA, Via Bonomea 265, I-34136 Trieste, Italy}
\affiliation{International Centre for Theoretical Physics (ICTP), P.O.Box 586, I-34014 Trieste, Italy}
\affiliation{CNR-IOM Democritos National Simulation Center, Via Bonomea 265, I-34136 Trieste, Italy}

\begin{abstract}
	We study Thouless pumping out of the adiabatic limit. Our findings show that despite its topological nature, this phenomenon is not {generically} robust to non-adiabatic effects. Indeed we find that the Floquet diagonal ensemble value of the pumped charge shows a deviation from the topologically
	quantized limit which is quadratic in the driving frequency {for a sudden switch-on of the driving}.
	This is reflected also in the charge pumped in a single period, which shows a non-analytic behaviour on top of an overall quadratic decrease.
	 {Exponentially small corrections are recovered only with a careful tailoring of the driving protocol.}
	We also discuss thermal effects and the experimental feasibility of observing such a deviation.
\end{abstract}
\maketitle

{\em Introduction.}
The quantization of the charge transported upon a cyclic adiabatic driving of a band insulating system, known as Thouless {topological} pumping, is a cornerstone of condensed matter physics~\cite{Thouless_PRB83}, recently experimentally realized in systems of ultracold atoms in optical lattices~\cite{Nakajima_Nphys16,Lohse_Nphys16}.
It laid the foundations of the field of charge pumping in mesoscopic systems~\cite{Altshuler_Sci99} and played a central role in the development of the modern theory of polarization~\cite{Kingsmith_PRB93,Ortiz_PRB94}.
Moreover, despite being a dynamical phenomenon, it is a conceptual key for understanding
many equilibrium properties related to {the topology of the bands in momentum space}. 
Most famously, the quantization of the Hall conductance in the Integer Quantum Hall effect (IQHE), through Laughlin's argument~\cite{Laughlin_PRB81,Kane_Book13}, can be seen as a Thouless {topological pump}. 
The quantization of the transported charge due to quantum topological effects crucially differentiates Thouless pumping from related phenomena. 
For example, parametric pumping~\cite{Note_Cohen} can be of geometric origin~
\cite{Brower98}, but is in general not characterized by a topological quantization. 
Furthermore, some types of parametric pumping, as ratchets~\cite{Schanz_PRE05} or piston-like pumps~\cite{Cohen_PRE05} ---
which share some formal analogies with Thouless pumping --- have a classical counterpart. 
On the contrary, quantum tunnelling effects are essential in making the charge quantization in Thouless pumping insensitive to a fine 
tuning of the model parameters~\cite{troyo_prl13}.
%
%
%
Having a topological nature, the quantization of the transported charge shows robustness to various factors, 
such as disorder or interactions~\cite{Niu_JPA84}.
Non-adiabatic effects are also believed to be unimportant --- exponentially small in the driving frequency $\omega$ \cite{Niu_PRL90,Shih_PRB94} ---
in analogy with the IQHE, where the Hall plateaus show corrections that are exponentially small in the longitudinal electric field~\cite{Klitzing_review}.
Theoretically, this follows from the fact that the quantized Chern number expression for the Hall conductivity, usually obtained through a Kubo formula 
in linear response, is valid at all orders in perturbation theory~\cite{Klein_CIMP90,Avron_JPA99}.

In this letter we study Thouless pumping 
out of the perfect adiabatic limit $\omega\to 0$.
In order to do that, we perform a careful Floquet analysis of a closed, clean, non-interacting system --- the driven Rice-Mele model --- in the thermodynamic limit.
By analyzing the charge pumped after many cycles when the system starts from the initial ground-state Slater determinant, 
{we find that for a suddenly switched-on driving}, 
the pumped charge shows a deviation from perfect quantization that is always \textit{polynomial} in the driving frequency $\omega$, contradicting the expected topological robustness. This quadratic deviation is present also after a finite number of pumping cycles, even if apparently hidden under a highly oscillatory non-analytic behaviour \cite{Avron_JPA99} in $\omega$.
An exponentially small deviation would be obtained only if one was able to prepare the system in a {specific} Floquet state, 
{which can be approximately obtained only with a suitable switch-on of the driving}.
We also discuss the effects of a thermal initial state.

{\em Model and Method.}
A paradigmatic model for 
 Thouless pumping 
is the driven Rice-Mele (RM) \cite{Rice_PRL82} model:
%
\begin{align} \label{RM-model:eqn}
\Ham_{\rm RM}(t) =
- & \sum_{j=1}^N \left( J_1(t) \, \opbdag{j}\opa{j} + J_2(t) \, \opadag{j+1}\opb{j} + {\rm H.c.} \right) \nonumber \\
+ & \sum_{j=1}^N \Delta(t)\left(   \opadag{j}\opa{j} - \opbdag{j}\opb{j} \right).
\end{align}
Here $\opadag{j}$ and $\opbdag{j}$ create a spinless fermion at cell $j$ in sublattice $A$ and $B$, respectively, and we assume a half-filling situation. 
This simple tight-binding model describes the physics of cold atoms experiments in some regimes~\cite{Nakajima_Nphys16,Lohse_Nphys16}.
The instantaneous spectrum becomes gapless for $J_1=J_2$ and $\Delta=0$, and a quantized adiabatic pumping 
is realized
when a closed path in the $(J_1-J_2,\Delta)$ parameter space encloses such a degeneracy point~\cite{xiao2010berry}.
In the following, we will parameterize $J_{\stackrel{1}{\scriptscriptstyle 2}} (t) = J_0  \pm \delta_0\cos(\varphi(t)) $, and $\Delta (t) = \Delta_0 \sin(\varphi(t))$.
{By choosing $\varphi(t) = \omega t$ we realize a sudden switch-on of the driving. (We will also discuss different choices of $\varphi(t)$.)}
We impose periodic boundary conditions (PBC), and use momentum $k$ {in the Brillouin Zone (BZ) $[-\frac{\pi}{a},\frac{\pi}{a})$}
to reduce the dynamics to $N$ independent $2$-dimensional Schr\"odinger problems,
which can be numerically integrated by a fourth-order Runge-Kutta method.

{\em Floquet theory of the Thouless pump.}
%
\begin{figure}
\begin{center} \includegraphics[width=0.49\textwidth]{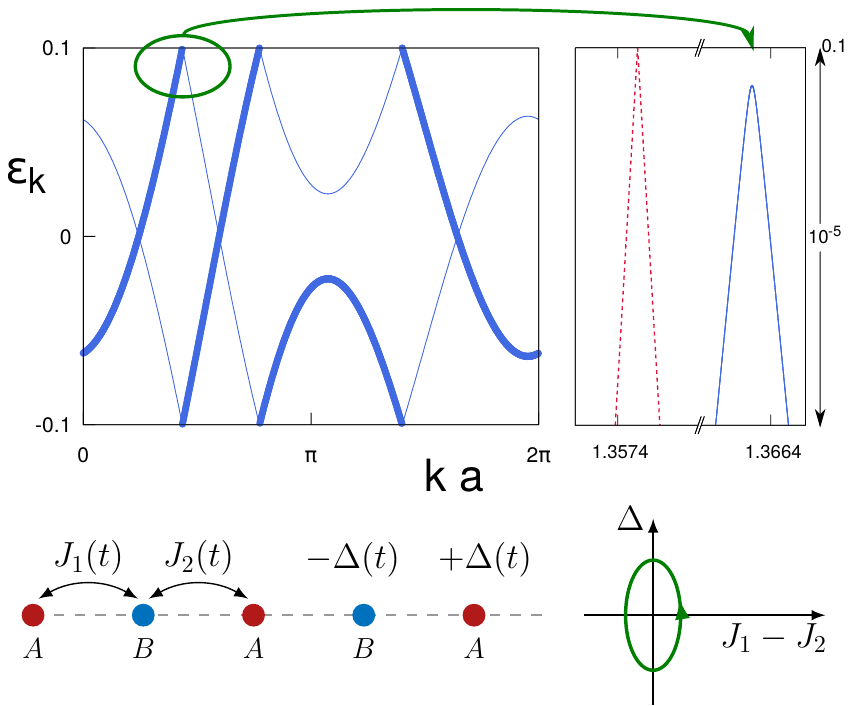}
	\end{center}
\caption{Top left: quasienergy spectrum of Rice-Mele model $\Delta_0 = 3 J_0$, $\delta_0 = J_0$, $\omega = 0.2J_0 /\hbar$.
The thick band is the {lowest-energy Floquet band $\varepsilon_{\LE,k}$}. Top right: (solid line) zoom of the previous figure close to the upper border of the FBZ around $ka = 1.3664$;
the dashed line denotes the quasienergies in the adiabatic limit  $\varepsilon_{\alpha,k}^{0}$.
Notice the gap of order $10^{-6}$.
Bottom: A cartoon of the Rice-Mele model (left) and a path in parameters space.}
\label{fig1}
\end{figure}
%
Given the time-periodicity of the Hamiltonian in a Thouless pump, {with period $\tau=2\pi/\omega$}, it is natural to employ a Floquet analysis \cite{Ferrari_IJMP98,Avron_JPA99,kitagawa2010topological,Shih_PRB94,Russomanno_PRB11}.
Because of the discrete time-translation invariance, there exists a basis of solutions of the time-dependent Schr\"{o}dinger equation, that are periodic up to a phase:
the Floquet states $\ket{\psi_\alpha (t)} = \nep^{-\frac{i}{\hbar} \varepsilon_{\alpha} t} \ket{\phi_\alpha (t)} $~\cite{sambe1973steady,Shirley_PR65}.
The $\tau$-periodic states $\ket{\phi_\alpha (t)}$ are the so-called Floquet modes and $\varepsilon_\alpha$ are the quasienergies:
they are defined modulo an integer number of $\hbar\omega=2\pi \hbar/\tau$, hence it is possible to restrict them to the first Floquet Brillouin zone (FBZ)  
$[-\hbar\omega/2,\hbar\omega/2)$.

In a PBC ring geometry, the total current operator $\hat{J}(t)$ is obtained as a derivative of $\hat{H}(t)$ with respect to a flux
$\Phi$ threading the ring, $\hat{J}=\partial_{\kappa} \hat{H}/\hbar$, where $\kappa = \frac{2\pi}{L} \frac{\Phi}{\Phi_0}$,
$L$ is the length of the system and $\Phi_0$ the flux quantum.
As a consequence, the charge pumped in one period $\tau$ by a single Floquet state $|\psi_{\alpha}(t)\rangle$
is ~\cite{Ferrari_IJMP98,Avron_JPA99,kitagawa2010topological,Shih_PRB94}
$Q_{\alpha}(\tau) = \frac{1}{L} \int_{0}^{\tau} \mathrm{d}t \, \bra{\psi_{\alpha}(t)} \hat{J}(t) \ket{\psi_{\alpha}(t)}= \frac{\tau}{\hbar L} \partial_{\kappa} \varepsilon_{\alpha}$.
%
For a  translationally-invariant system, each completely filled Floquet-Bloch band with (single-particle) quasienergy dispersion $\varepsilon_{\alpha,k}$
would contribute to the charge pumped (in the thermodynamic limit $L\to \infty$) as
\begin{equation} \label{eqn:Q}
Q_{\alpha}(\tau) = \frac{1}{\hbar \omega} \int_{-\frac{\pi}{a}}^{+\frac{\pi}{a}} \!\! \mathrm{d} k \, \frac{\partial \varepsilon_{\alpha,k}}{\partial k} \,
\end{equation}
where we have replaced the $\kappa$-derivative with a $k$-derivative, since $\varepsilon_{\alpha,k}$ depends on $k+\kappa$.
Thus, if $\varepsilon_{\alpha,k}$ wraps around the FBZ in a continuous way as a function of $k$, $Q_{\alpha}(\tau)$ is equivalent to the {\em winding number} of the band, i.e.,
the number $n$ of times $\varepsilon_{\alpha,k}$ goes around the FBZ, 
$\varepsilon_{\alpha,+\frac{\pi}{a}} - \varepsilon_{\alpha,-\frac{\pi}{a}} = n \hbar\omega $,
and $Q_{\alpha}(\tau)$ is therefore quantized{: $Q_{\alpha}(\tau)=n$}.
This is what happens in the extreme adiabatic limit $\omega\to 0$: if $|\Psi_{\alpha}(t)\rangle$ is a Slater determinant made up of the instantaneous Hamiltonian Bloch eigenstates 
$e^{ikx} u_{\alpha,k}(x,t)$ belonging to a filled band $E_{\alpha,k}(t)$, the adiabatic theorem guarantees that such a state returns onto itself after a period $\tau$,
$|\Psi_{\alpha}(\tau)\rangle = e^{i\sum_{k}^{\mathrm{BZ}}(\gamma_{\alpha,k}-\theta_{\alpha,k})} |\Psi_{\alpha}(0)\rangle$,
by acquiring a geometric (Berry) phase
$\gamma_{\alpha,k} =  \int_{0}^{\tau} \mathrm{d}t \, i\bra{u_{\alpha,k}} \partial_t u_{\alpha,k} \rangle$
and a dynamical one $\theta_{\alpha,k} = \int_{0}^{\tau} \mathrm{d}t \, E_{\alpha,k}(t) / \hbar$.
This in turn implies that $|\Psi_{\alpha}(t)\rangle$ is a Floquet state with quasienergy $\varepsilon_{\alpha,k}^0 = \hbar(-\gamma_{\alpha,k} + \theta_{\alpha,k})/\tau$.
Substituting in~
{Eq.~\eqref{eqn:Q}}, only the geometric phase {survives, leading to the} Thouless' formula~\cite{Thouless_PRB83}
\begin{equation}
Q_{\alpha}(\tau) =  \int_{-\frac{\pi}{a}}^{+\frac{\pi}{a}} \! \frac{\mathrm{d}k}{2\pi} \int_{0}^{\tau} \!\! \mathrm{d}t \,
i(\bra{\partial_{k}u_{\alpha,k}} \partial_{t} u_{\alpha,k}\rangle - {\mathrm{c.c.}}) \,
\end{equation}
identifying the pumped charge with a Chern number \cite{Avron_JPA99}.

Let us see what happens away from the adiabatic limit $\omega\to 0$.
We consider 
a lattice model with a finite number of bands, such as Eq.~\eqref{RM-model:eqn}.
The sum of the winding numbers of all bands will be zero, since the sum of Chern numbers of a finite-dimensional Hamiltonian must be zero \cite{Avron_JPA99}.
This fact, as noticed in Ref.~\cite{Avron_JPA99}, implies that the quasienergy spectrum must contain some crossings if at least one quasienergy band has non vanishing winding number.
These crossings, however, are not stable \cite{Avron_JPA99}: according to Wigner and von Neumann~\cite{wigner_1929}, a true crossing requires,
for the present case of a complex $2\times 2$ unitary operator,
the tuning of at least three real parameters, while the quasienergy spectrum depends only on two, $\tau$ and $k$.
Hence one expects, {\em generically}, that crossings turn into avoided crossings with opening of gaps for any finite $\tau$ --- in the present case at the border of the FBZ ---
implying a deviation from perfect quantization of the pumped charge for the Floquet band under consideration.
%

\begin{figure}
\includegraphics[width=0.48\textwidth]{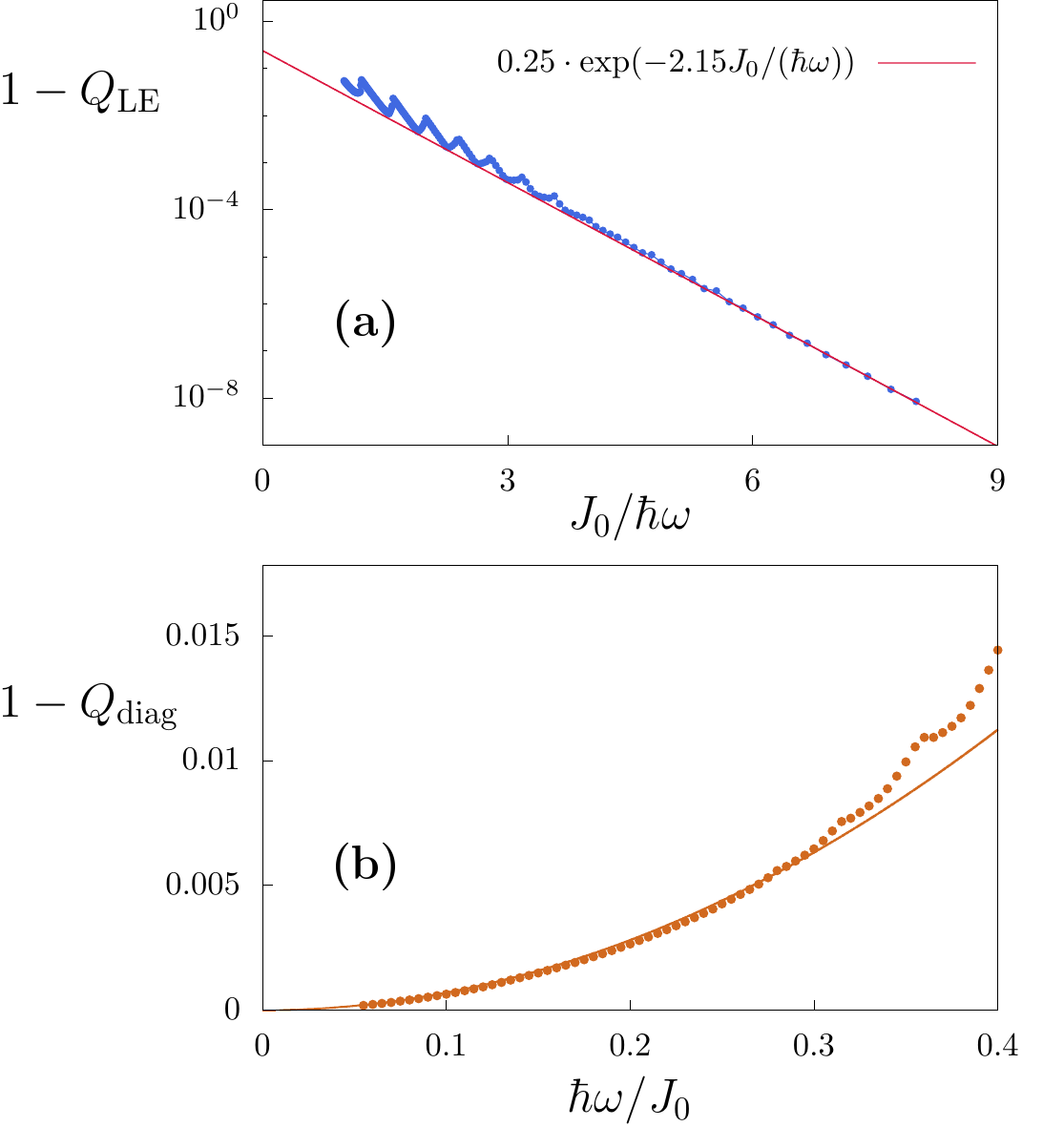}
\caption{(a) The deviation from $1$ of the charge pumped by the lowest-energy Floquet band (proportional to the sum of gaps) as a function of $1/\hbar\omega$,
{with its exponential fit (red solid straight line)}.
(b) Deviations from $1$ of the diagonal pumped charge $Q_{\rm diag}$ in the RM model {for a sudden switch-on of the driving}. 
The smooth curve $\frac{9}{128}(\hbar\omega/J_0)^2 $ is obtained from 
Eq.~\eqref{npert2:eqn}.
The model parameters are $\Delta_0 = 3 J_0$, $\delta_0 = J_0$.
}
\label{fig2}
\end{figure}

To better understand this point, let us focus on the Floquet-Bloch band whose pumped charge is {closest to the integer value of the adiabatic limit}. 
This band can be constructed~\cite{Shih_PRB94} by choosing, for each $k$, the Floquet mode with {(period-averaged) lowest-energy} expectation 
$\ket{\phi_{{\LE}, k}(t)}$.
In the left panel of Fig.~\ref{fig1} we show a typical quasienergy spectrum 
of the RM model:
the bold line denotes the lowest-energy Floquet band.
In the right panel we zoom in the region around an avoided crossing (solid line),
comparing with the perfect crossing occurring for the adiabatic {approximation}  $\varepsilon_{\alpha,k}^0$ (dashed line):
the visible gap is exponentially small in $1/\omega$ --- as it happens for all the gaps that open at such avoided crossings \cite{Lindner_PRX17}.
This implies that the lowest-energy Floquet band does {\em not} wrap continuously around the FBZ. 
Accordingly, its pumped charge $Q_{\LE} = (\hbar \omega)^{-1}  \int_{-\frac{\pi}{a}}^{+\frac{\pi}{a}} \mathrm{d}k \, \partial_k \varepsilon_{{\LE},k}$
deviates from an integer by terms proportional to the sum of the gaps when $\omega>0$.
This deviation is therefore exponentially small in $1/\omega$ (a similar result was found in Ref.~\cite{Shih_PRB94} for a different model),
see Fig.~\ref{fig2}(a).
Summarizing, if we were able to prepare an initial state  coinciding with the lowest-energy Floquet band, the deviation from perfect quantization would be exponentially small.
Nevertheless, in any real situation, the initial state $\ket{\Psi(0)}$ of the system is not a Floquet state:
a more realistic starting point would be to assume that $\ket{\Psi(0)}$ is the {\em ground state} of the initial Hamiltonian $\hat{H}(0)$ before pumping is started.
Whichever the initial state, any local observable attains, upon periodic driving and in the thermodynamic limit, a periodic steady state with the same periodicity as the driving~\cite{Russomanno_PRL12}.
This asymptotic regime is described by the Floquet diagonal density matrix ~\cite{Russomanno_PRL12,Lazarides_PRL14};
in an integrable system this density matrix is not thermal, but is
given by a generalized Gibbs ensemble. 
Let us denote by $Q(m\tau)$ the total charge pumped in the first $m$ periods starting from the initial ground state $|\Psi(0)\rangle$ of $\hat{H}(0)$.
The {\em asymptotic} charge pumped in a single cycle, obtained from the infinite time limit, is given by the Floquet diagonal ensemble~\cite{Note_Supp}: \nocite{Avron_CMP87}
%
\begin{equation} \label{eq:ChargeDiag}
Q_{\rm diag} \equiv \lim_{m\to\infty} \frac{Q(m\tau)}{m} =
\frac{1}{\hbar \omega} \sum_{\alpha} \int_{-\frac{\pi}{a}}^{+\frac{\pi}{a}} \! \mathrm{d}k \; n_{\alpha,k}
\frac{\partial \varepsilon_{\alpha,k}}{\partial k} \,
\end{equation}
where $n_{\alpha,k}=\bra{\Psi(0)} \opfdag{\alpha,k} \opf{\alpha,k}\ket{\Psi(0)}$ is the initial ground-state occupation of the Floquet-Bloch $(\alpha,k)$-mode,
with $\opfdag{\alpha,k}\ket{0}= \ket{\phi_{\alpha,k}(0)}$.
The occupations $n_{\alpha,k}$ {can} give rise to a stronger deviation from quantization than the gaps, as we are now showing.

\begin{figure} \centering
 \includegraphics[width=0.48\textwidth]{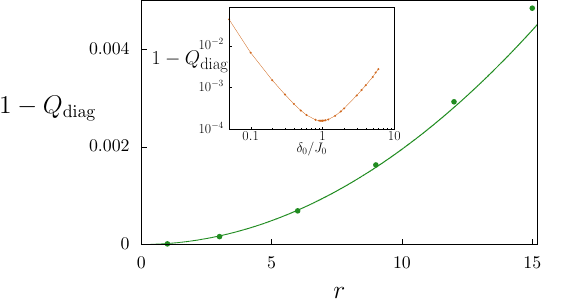}
	\caption{Deviations from {the integer value} of the diagonal pumped charge $Q_{\rm diag}$ in the RM model versus the aspect ratio
		$r=\Delta_0/\delta_0$ of the driving ellipse for $\delta_0=J_0$, 
		{for a sudden switch-on of the driving with frequency $\omega = 0.05 J_0/\hbar$}. 
		The smooth curve is $\frac{1}{128} (\hbar\omega/J_0)^2 \, r^2$ and it was calculated with the first order perturbative calculation of the main text.
		In the inset, the deviation from $1$ of $Q_{\rm diag}$ for a fixed  $r=3$ as a function of $\delta_0$:
		The minimum sits at $\delta_0=J_0$.
	}
	\label{fig3}
\end{figure}

{\em Results.}
%
%
In Fig.~\ref{fig2}(b) we plot the diagonal pumped charge $Q_{\rm diag}$, calculated from Eq.~\eqref{eq:ChargeDiag},
for the RM model \eqref{RM-model:eqn} as a function of the driving frequency $\omega$, close to the adiabatic limit $\omega\to 0$,
for the same parameters as Fig.~\ref{fig1}. {The driving is suddenly switched on: $\varphi(t)=0$ when $t<0$ and $\varphi(t)=\omega t$ for $t\geq 0$, 
as realized in the experimental setting of Ref.~\onlinecite{Nakajima_Nphys16}.}
The numerically determined points show a clear \textit{quadratic} deviation with $\omega$ from the fully adiabatic integer value $1$.
We now show that this power-law deviation essentially originates from the Floquet bands occupations $n_{\alpha,k}$. 
To understand this point, consider the lowest-energy Floquet band $\varepsilon_{{\LE},k}$, and the associated occupations
$n_{{\LE},k}=\bra{\Psi(0)} \opfdag{{\LE},k} \opf{{\LE},k}\ket{\Psi(0)}$.
One can develop a perturbation theory in $\omega$ for the Floquet modes, along the lines of Ref.~\cite{Rigolin_PRA08}, to show that for our model
\begin{align} \label{npert1:eqn}
n_{{\LE},k} = 1 - \left|\dfrac{\hbar\omega \bra{u_{1,k}(0)} \partial_s u_{0,k}(0) \rangle}{E_{1,k} (0) - E_{0,k}(0)} \right|^2 + \mathcal{O}(\omega^3).
\end{align}
Here $s=t/\tau$ is a rescaled time, while $E_{\alpha,k}(t)$ and $\ket{u_{\alpha,k}(t)}$, with $\alpha = 0,1$, are the energy and the periodic part of the two instantaneous Bloch eigenfunctions. The analytic calculation is simple:
\begin{equation} \label{npert2:eqn}
n_{{\LE},k} = 1- \frac{1}{64} \left( \frac{\hbar\omega \Delta_0 }{J_0^2 + \delta^2_0  +(J_0^2-\delta^2_0)\cos(ka)} \right)^2 + \; ...
\end{equation}
leading 
to quadratic corrections to $Q_{\rm diag}$.
When $\delta_0=J_0$, Eq.~\eqref{npert2:eqn} predicts that $n_{{\LE},k}$ is $k$-independent and can be taken out of the integral in Eq.~\eqref{eq:ChargeDiag}:
we can calculate the charge deviation given by $\frac{1}{128}(\hbar\omega\Delta_0/J^2_0)^2$, which perfectly fits the numerical data points,
as shown in Fig.~\ref{fig2}(a) for $\Delta_0 = 3 J_0$.
It is apparent from Eq.~\eqref{npert2:eqn} that the non-adiabatic corrections can be more or less pronounced, depending on the parameters of the driving $(\Delta_0, \delta_0)$.
Figure~\ref{fig3} illustrates how the deviation from quantization for a fixed value of frequency, $\omega = 0.05 J_0/\hbar$, depends on $(\Delta_0, \delta_0)$.
In the main plot, we fix the value $\delta_0 = J_0$ as before and we vary the dimensionless ratio $r= \Delta_0 / \delta_0$.
According to Eq.~\eqref{npert2:eqn}, the deviation increases with $r$ as $\frac{\hbar^2\omega^2}{128 J_0^2}r^2$.
The agreement with the numerical data for this choice of $\omega$ is very good up to $r\lesssim 12$, where higher orders of the perturbation theory~\eqref{npert1:eqn} become relevant.
Conversely, in the inset of Fig.~\ref{fig3} we fix $r= \Delta_0/\delta_0 = 3$ and we consider the dependence on $\delta_0/J_0$: we see that the deviation shows a minimum around $\delta_0=J_0$. Summarizing, the deviation should be more noticeable for paths in parameters space flattened on the $\Delta$ axis and far from 
 $\delta_0 = J_0$.
Note that a small $\delta_0/J_0$ corresponds to a weak pumping regime: quite surprisingly, this seems to  imply a stronger non-adiabaticity. 
The possibility of controlling Hamiltonian parameters in ultracold atoms experiments makes the detection of these non-adiabatic effects likely feasible.

{It is however possible to devise driving schemes that lead to a better filling of the lowest-energy Floquet band. 
The Floquet adiabatic theorem~\cite{Young_JMP70,breuer1989adiabatic,Eckardt_PRL05,Russomanno_EPL16} suggests that a sufficiently smooth variation of the 
instantaneous driving frequency $\omega(t) = \dot{\varphi}(t)$ would lead to a much smaller deviation of the population $n_{\LE, k}$, and hence of $Q_{\rm diag}$, 
from an integer value. 
This is what a detailed analysis of these issues, presented elsewhere~\cite{Wauters_unpub},  finds. 
Incidentally, a smoother switch-on of the periodic driving is what the experimental realization of Ref.~\onlinecite{Lohse_Nphys16} adopts. 
} 

\begin{figure}	
        \includegraphics[width=0.48\textwidth]{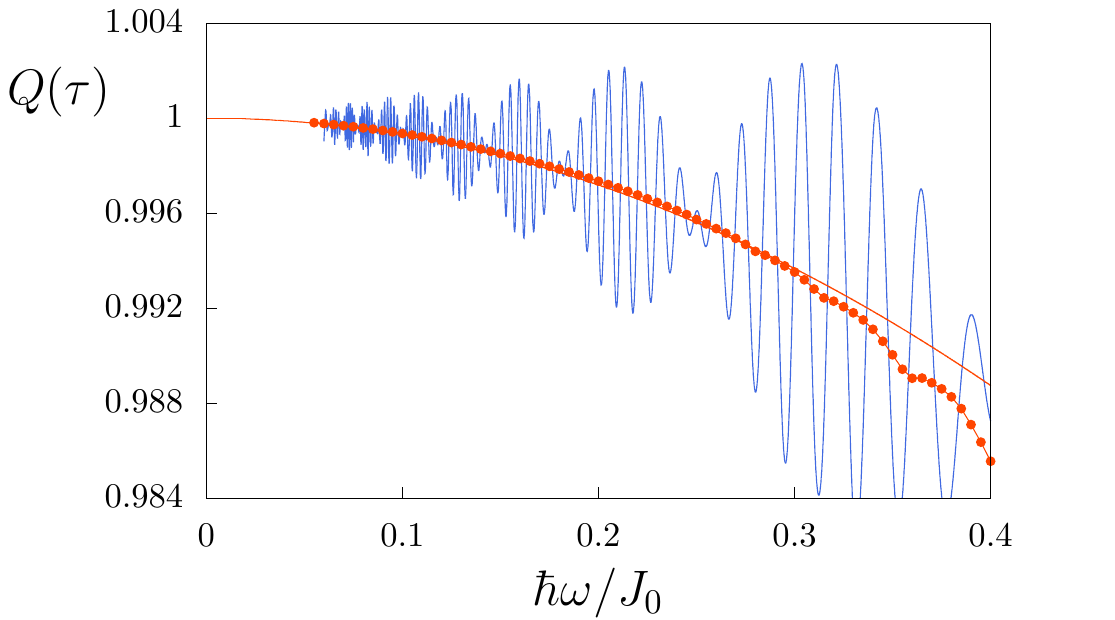}
	\caption{
	The charge pumped after the first period, $Q(\tau)$, {as a function of the frequency $\omega$, 
	                  for the RM model with a suddenly switched-on driving (smooth blue line).} 
	The {red dotted line} is the corresponding diagonal ensemble value $Q_{\rm diag}$, reported in Fig.~\ref{fig2}(b).
	The model parameters are $\Delta_0 = 3 J_0$, $\delta_0 = J_0$.}
	\label{fig4}
\end{figure}

We now address the issue of non-adiabatic deviation for a {\em finite} number of pumping cycles.
Diagonal expectation values  are indeed attained after some transient and become exact only after an infinite number of pumping cycles.
In Fig.~\ref{fig4} we plot the charge pumped after a single cycle, $Q(\tau)$, as a function of $\omega$: we see that $Q(\tau)$ exhibits remarkable beating-like oscillations,
on top of the overall quadratic decrease of $Q_{\rm diag}$, which become faster and faster as $\omega\to 0$.
The theoretical prediction, according to a theorem of Ref.~\cite{Avron_JPA99}, is that the finite-time pumped charge must have an essential singularity in $\omega=0$.
The behaviour that we find is indeed compatible with the presence of non-analyticities, possibly of the kind of $\sin(c/\omega)$.

An alternative source of deviation from perfect quantization is finite temperature. 
In 
ultracold atoms experiments, it is reasonable to consider the dynamics to be coherent even if the initial state is a thermal density matrix $\hat{\rho}_\mathsmaller{T}$ at temperature $T$. 
The zero-temperature unitary evolution results of Eq.~\eqref{eq:ChargeDiag} get modified only through the replacement of {the} occupations
$n_{\alpha,k}$ with thermal ones $n^T_{\alpha,k} = \Tr \left( {\hat{\rho}}_{\mathsmaller{T}}\opfdag{\alpha,k} \opf{\alpha,k}\right)$.
The final result is 
\begin{equation}
{
n^T_{\alpha,k}= \tanh(-\beta E_{0,k}(0)) n_{\alpha,k}  + \frac{e^{\beta E_{0,k}(0)}}{2 \cosh(-\beta E_{0,k}(0))} \;,
}
\end{equation}
where we used that $E_{0,k} = - E_{1,k}$ in the RM model. 
Thus, the $\omega^2$ behaviour of the deviation is preserved, but can be hidden by thermal effects.
They are exponentially small when the temperature $T$ is much smaller than the initial gap. 
For the specific choice used before, $\delta_0=J_0$, thermal corrections only amount to multiplying the $T=0$ result by a factor 
$\tanh(\beta E_{1,k}(0))$, which turns out to be $k$-independent. 
In this case $E_{1,k} (0)=2J_0$ and thermal effects start to compete with the non-adiabatic ones only when $T$ is of the order of the gap, $k_BT  \approx J_0$.

{\em Conclusions.}
We have studied what happens to the quantization of the Thouless pumped charge out of the perfect adiabatic limit.
Within a Floquet framework, we have found that this transport phenomenon is {in general} not robust  to non-adiabatic effects despite its topological nature.
{When the driving is switched on suddenly, $\phi(t)=\omega t$, or too fast~\cite{Wauters_unpub}, we see that the long-time asymptotic value} of the pumped charge 
deviates from the quantized value in a polynomial fashion, i.e. quadratically in the driving frequency.
This observation is, we believe, model-independent, since it requires only that the initial state is the ground state (or any other eigenstate) of the initial Hamiltonian.
The fact that a topologically robust property  {can be} ruined by the occupation factors of the Floquet bands is in line with  what was found in a resonantly driven
graphene layer \cite{Privitera_PRB16}.
Our findings 
should be in principle observable in ultracold atoms experiments 
(for instance with the methods used in~\cite{Lohse_Nphys16,Nakajima_Nphys16}).
Perspectives of future work include the study of the pumped charge in the pre-thermal regime of a 
non-integrable system~\cite{Lindner_PRX17} and the analysis of 
disorder, especially in connection with the stabilization of charge pumping in a many-body localized system. 
{Another important point will be understanding the switch-on time-scale marking the crossover between power-law and exponentially small deviations from quantized pumping.}

We thank J. Avron, M.V. Berry, E. Berg, I. Carusotto, M. Dalmonte, R. Fazio, F. Romeo and M.M. Wauters for discussions.
GES acknowledges support by the EU FP7 under ERC-MODPHYSFRICT, Grant Agreement No. 320796. 
AR acknowledges financial support from EU through project QUIC
(under Grant Agreement No. 641122) and from ``Progetti interni - Scuola Normale Superiore''.


\begin{thebibliography}{39}%
	\makeatletter
	\providecommand \@ifxundefined [1]{%
		\@ifx{#1\undefined}
	}%
	\providecommand \@ifnum [1]{%
		\ifnum #1\expandafter \@firstoftwo
		\else \expandafter \@secondoftwo
		\fi
	}%
	\providecommand \@ifx [1]{%
		\ifx #1\expandafter \@firstoftwo
		\else \expandafter \@secondoftwo
		\fi
	}%
	\providecommand \natexlab [1]{#1}%
	\providecommand \enquote  [1]{``#1''}%
	\providecommand \bibnamefont  [1]{#1}%
	\providecommand \bibfnamefont [1]{#1}%
	\providecommand \citenamefont [1]{#1}%
	\providecommand \href@noop [0]{\@secondoftwo}%
	\providecommand \href [0]{\begingroup \@sanitize@url \@href}%
	\providecommand \@href[1]{\@@startlink{#1}\@@href}%
	\providecommand \@@href[1]{\endgroup#1\@@endlink}%
	\providecommand \@sanitize@url [0]{\catcode `\\12\catcode `\$12\catcode
		`\&12\catcode `\#12\catcode `\^12\catcode `\_12\catcode `\%12\relax}%
	\providecommand \@@startlink[1]{}%
	\providecommand \@@endlink[0]{}%
	\providecommand \url  [0]{\begingroup\@sanitize@url \@url }%
	\providecommand \@url [1]{\endgroup\@href {#1}{\urlprefix }}%
	\providecommand \urlprefix  [0]{URL }%
	\providecommand \Eprint [0]{\href }%
	\providecommand \doibase [0]{http://dx.doi.org/}%
	\providecommand \selectlanguage [0]{\@gobble}%
	\providecommand \bibinfo  [0]{\@secondoftwo}%
	\providecommand \bibfield  [0]{\@secondoftwo}%
	\providecommand \translation [1]{[#1]}%
	\providecommand \BibitemOpen [0]{}%
	\providecommand \bibitemStop [0]{}%
	\providecommand \bibitemNoStop [0]{.\EOS\space}%
	\providecommand \EOS [0]{\spacefactor3000\relax}%
	\providecommand \BibitemShut  [1]{\csname bibitem#1\endcsname}%
	\let\auto@bib@innerbib\@empty
	\bibitem [{\citenamefont {Thouless}(1983)}]{Thouless_PRB83}%
	\BibitemOpen
	\bibfield  {author} {\bibinfo {author} {\bibfnamefont {D.~J.}\ \bibnamefont
			{Thouless}},\ }\href@noop {} {\bibfield  {journal} {\bibinfo  {journal}
			{Phys. Rev. B}\ }\textbf {\bibinfo {volume} {27}},\ \bibinfo {pages} {6083}
		(\bibinfo {year} {1983})}\BibitemShut {NoStop}%
	\bibitem [{\citenamefont {Nakajima}\ \emph {et~al.}(2016)\citenamefont
		{Nakajima}, \citenamefont {Tomita}, \citenamefont {Taie}, \citenamefont
		{Ichinose}, \citenamefont {Ozawa}, \citenamefont {Wang}, \citenamefont
		{Troyer},\ and\ \citenamefont {Takahashi}}]{Nakajima_Nphys16}%
	\BibitemOpen
	\bibfield  {author} {\bibinfo {author} {\bibfnamefont {S.}~\bibnamefont
			{Nakajima}}, \bibinfo {author} {\bibfnamefont {T.}~\bibnamefont {Tomita}},
		\bibinfo {author} {\bibfnamefont {S.}~\bibnamefont {Taie}}, \bibinfo {author}
		{\bibfnamefont {T.}~\bibnamefont {Ichinose}}, \bibinfo {author}
		{\bibfnamefont {H.}~\bibnamefont {Ozawa}}, \bibinfo {author} {\bibfnamefont
			{L.}~\bibnamefont {Wang}}, \bibinfo {author} {\bibfnamefont {M.}~\bibnamefont
			{Troyer}}, \ and\ \bibinfo {author} {\bibfnamefont {Y.}~\bibnamefont
			{Takahashi}},\ }\href@noop {} {\bibfield  {journal} {\bibinfo  {journal}
			{Nat. Phys.}\ }\textbf {\bibinfo {volume} {12}},\ \bibinfo {pages} {296}
		(\bibinfo {year} {2016})}\BibitemShut {NoStop}%
	\bibitem [{\citenamefont {Lohse}\ \emph {et~al.}(2016)\citenamefont {Lohse},
		\citenamefont {Schweizer}, \citenamefont {Zilberberg}, \citenamefont
		{Aidelsburger},\ and\ \citenamefont {Bloch}}]{Lohse_Nphys16}%
	\BibitemOpen
	\bibfield  {author} {\bibinfo {author} {\bibfnamefont {M.}~\bibnamefont
			{Lohse}}, \bibinfo {author} {\bibfnamefont {C.}~\bibnamefont {Schweizer}},
		\bibinfo {author} {\bibfnamefont {O.}~\bibnamefont {Zilberberg}}, \bibinfo
		{author} {\bibfnamefont {M.}~\bibnamefont {Aidelsburger}}, \ and\ \bibinfo
		{author} {\bibfnamefont {I.}~\bibnamefont {Bloch}},\ }\href@noop {}
	{\bibfield  {journal} {\bibinfo  {journal} {Nat. Phys.}\ }\textbf {\bibinfo
			{volume} {12}},\ \bibinfo {pages} {350} (\bibinfo {year} {2016})}\BibitemShut
	{NoStop}%
	\bibitem [{\citenamefont {Altshuler}\ and\ \citenamefont
		{Glazman}(1999)}]{Altshuler_Sci99}%
	\BibitemOpen
	\bibfield  {author} {\bibinfo {author} {\bibfnamefont {B.~L.}\ \bibnamefont
			{Altshuler}}\ and\ \bibinfo {author} {\bibfnamefont {I.}~\bibnamefont
			{Glazman}},\ }\href {\doibase 10.1126/science.283.5409.1864} {\bibfield
		{journal} {\bibinfo  {journal} {Science}\ }\textbf {\bibinfo {volume}
			{283}},\ \bibinfo {pages} {1864} (\bibinfo {year} {1999})}\BibitemShut
	{NoStop}%
	\bibitem [{\citenamefont {King-Smith}\ and\ \citenamefont
		{Vanderbilt}(1993)}]{Kingsmith_PRB93}%
	\BibitemOpen
	\bibfield  {author} {\bibinfo {author} {\bibfnamefont {R.}~\bibnamefont
			{King-Smith}}\ and\ \bibinfo {author} {\bibfnamefont {D.}~\bibnamefont
			{Vanderbilt}},\ }\href@noop {} {\bibfield  {journal} {\bibinfo  {journal}
			{Phys. Rev. B}\ }\textbf {\bibinfo {volume} {47}},\ \bibinfo {pages} {1651}
		(\bibinfo {year} {1993})}\BibitemShut {NoStop}%
	\bibitem [{\citenamefont {Ortiz}\ and\ \citenamefont
		{Martin}(1994)}]{Ortiz_PRB94}%
	\BibitemOpen
	\bibfield  {author} {\bibinfo {author} {\bibfnamefont {G.}~\bibnamefont
			{Ortiz}}\ and\ \bibinfo {author} {\bibfnamefont {R.~M.}\ \bibnamefont
			{Martin}},\ }\href@noop {} {\bibfield  {journal} {\bibinfo  {journal} {Phys.
				Rev. B}\ }\textbf {\bibinfo {volume} {49}},\ \bibinfo {pages} {14202}
		(\bibinfo {year} {1994})}\BibitemShut {NoStop}%
	\bibitem [{\citenamefont {Laughlin}(1981)}]{Laughlin_PRB81}%
	\BibitemOpen
	\bibfield  {author} {\bibinfo {author} {\bibfnamefont {R.~B.}\ \bibnamefont
			{Laughlin}},\ }\href@noop {} {\bibfield  {journal} {\bibinfo  {journal}
			{Phys. Rev. B}\ }\textbf {\bibinfo {volume} {23}},\ \bibinfo {pages} {5632}
		(\bibinfo {year} {1981})}\BibitemShut {NoStop}%
	\bibitem [{\citenamefont {Kane}(2013)}]{Kane_Book13}%
	\BibitemOpen
	\bibfield  {author} {\bibinfo {author} {\bibfnamefont {C.~L.}\ \bibnamefont
			{Kane}},\ }in\ \href@noop {} {\emph {\bibinfo {booktitle} {Topological
				Insulators}}},\ Vol.~\bibinfo {volume} {6},\ \bibinfo {editor} {edited by\
		\bibinfo {editor} {\bibfnamefont {M.}~\bibnamefont {Franz}}\ and\ \bibinfo
		{editor} {\bibfnamefont {L.~W.}\ \bibnamefont {Molenkamp}}}\ (\bibinfo
	{publisher} {Elsevier},\ \bibinfo {year} {2013})\ p.~\bibinfo {pages}
	{3}\BibitemShut {NoStop}%
	\bibitem [{Note_Cohen({\natexlab{a}})}]{Note_Cohen}%
	\BibitemOpen
	\bibinfo {note} {Thouless pumping differs from the most common definition of
		parametric pumping, which refers to a model device in which a compact region
		in space (scattering region) is connected ballistically to two external
		asymptotic regions (reservoir). While Thouless pumping is purely
		quantum~\cite{troyo_prl13}, parametric pumping can occur both in the
		classical and in quantum case (for instance in
		ratchets~\cite{Schanz_PRE05}).}\BibitemShut {Stop}%
	\bibitem [{\citenamefont {Brower}(1998)}]{Brower98}%
	\BibitemOpen
	\bibfield  {author} {\bibinfo {author} {\bibfnamefont {P.~W.}\ \bibnamefont
			{Brower}},\ }\href@noop {} {\bibfield  {journal} {\bibinfo  {journal} {Phys.
				Rev. B}\ }\textbf {\bibinfo {volume} {58}},\ \bibinfo {pages} {R10135}
		(\bibinfo {year} {1998})}\BibitemShut {NoStop}%
	\bibitem [{\citenamefont {Schanz}\ \emph {et~al.}(2005)\citenamefont {Schanz},
		\citenamefont {Dittrich},\ and\ \citenamefont {Ketzmerick}}]{Schanz_PRE05}%
	\BibitemOpen
	\bibfield  {author} {\bibinfo {author} {\bibfnamefont {H.}~\bibnamefont
			{Schanz}}, \bibinfo {author} {\bibfnamefont {T.}~\bibnamefont {Dittrich}}, \
		and\ \bibinfo {author} {\bibfnamefont {R.}~\bibnamefont {Ketzmerick}},\
	}\href {\doibase 10.1103/PhysRevE.71.026228} {\bibfield  {journal} {\bibinfo
		{journal} {Phys. Rev. E}\ }\textbf {\bibinfo {volume} {71}},\ \bibinfo
	{pages} {026228} (\bibinfo {year} {2005})}\BibitemShut {NoStop}%
\bibitem [{\citenamefont {Cohen}\ \emph {et~al.}(2005)\citenamefont {Cohen},
	\citenamefont {Kottos},\ and\ \citenamefont {Schanz}}]{Cohen_PRE05}%
\BibitemOpen
\bibfield  {author} {\bibinfo {author} {\bibfnamefont {D.}~\bibnamefont
		{Cohen}}, \bibinfo {author} {\bibfnamefont {T.}~\bibnamefont {Kottos}}, \
	and\ \bibinfo {author} {\bibfnamefont {H.}~\bibnamefont {Schanz}},\ }\href
{\doibase 10.1103/PhysRevE.71.035202} {\bibfield  {journal} {\bibinfo
		{journal} {Phys. Rev. E}\ }\textbf {\bibinfo {volume} {71}},\ \bibinfo
	{pages} {035202} (\bibinfo {year} {2005})}\BibitemShut {NoStop}%
\bibitem [{\citenamefont {Wang}\ \emph {et~al.}(2013)\citenamefont {Wang},
	\citenamefont {Troyer},\ and\ \citenamefont {Dai}}]{troyo_prl13}%
\BibitemOpen
\bibfield  {author} {\bibinfo {author} {\bibfnamefont {L.}~\bibnamefont
		{Wang}}, \bibinfo {author} {\bibfnamefont {M.}~\bibnamefont {Troyer}}, \ and\
	\bibinfo {author} {\bibfnamefont {X.}~\bibnamefont {Dai}},\ }\href@noop {}
{\bibfield  {journal} {\bibinfo  {journal} {Phys. Rev. Lett.}\ }\textbf
	{\bibinfo {volume} {111}},\ \bibinfo {pages} {026802} (\bibinfo {year}
	{2013})}\BibitemShut {NoStop}%
\bibitem [{\citenamefont {Niu}\ and\ \citenamefont
	{Thouless}(1984)}]{Niu_JPA84}%
\BibitemOpen
\bibfield  {author} {\bibinfo {author} {\bibfnamefont {Q.}~\bibnamefont
		{Niu}}\ and\ \bibinfo {author} {\bibfnamefont {D.}~\bibnamefont {Thouless}},\
}\href@noop {} {\bibfield  {journal} {\bibinfo  {journal} {J. Phys. A-Math.
		Gen.}\ }\textbf {\bibinfo {volume} {17}},\ \bibinfo {pages} {2453} (\bibinfo
{year} {1984})}\BibitemShut {NoStop}%
\bibitem [{\citenamefont {Niu}(1990)}]{Niu_PRL90}%
\BibitemOpen
\bibfield  {author} {\bibinfo {author} {\bibfnamefont {Q.}~\bibnamefont
		{Niu}},\ }\href@noop {} {\bibfield  {journal} {\bibinfo  {journal} {Phys.
			Rev. Lett.}\ }\textbf {\bibinfo {volume} {64}},\ \bibinfo {pages} {1812}
	(\bibinfo {year} {1990})}\BibitemShut {NoStop}%
\bibitem [{\citenamefont {Shih}\ and\ \citenamefont {Niu}(1994)}]{Shih_PRB94}%
\BibitemOpen
\bibfield  {author} {\bibinfo {author} {\bibfnamefont {W.-K.}\ \bibnamefont
		{Shih}}\ and\ \bibinfo {author} {\bibfnamefont {Q.}~\bibnamefont {Niu}},\
}\href@noop {} {\bibfield  {journal} {\bibinfo  {journal} {Phys. Rev. B}\
}\textbf {\bibinfo {volume} {50}},\ \bibinfo {pages} {11902} (\bibinfo {year}
{1994})}\BibitemShut {NoStop}%
\bibitem [{\citenamefont {von Klitzing}(1986)}]{Klitzing_review}%
\BibitemOpen
\bibfield  {author} {\bibinfo {author} {\bibfnamefont {K.}~\bibnamefont {von
			Klitzing}},\ }\href@noop {} {\bibfield  {journal} {\bibinfo  {journal} {Rev.
			Mod. Phys.}\ }\textbf {\bibinfo {volume} {58}},\ \bibinfo {pages} {519}
	(\bibinfo {year} {1986})}\BibitemShut {NoStop}%
\bibitem [{\citenamefont {Klein}\ and\ \citenamefont
	{Seiler}(1990)}]{Klein_CIMP90}%
\BibitemOpen
\bibfield  {author} {\bibinfo {author} {\bibfnamefont {M.}~\bibnamefont
		{Klein}}\ and\ \bibinfo {author} {\bibfnamefont {R.}~\bibnamefont {Seiler}},\
}\href@noop {} {\bibfield  {journal} {\bibinfo  {journal} {Comm. Math.
		Phys.}\ }\textbf {\bibinfo {volume} {128}},\ \bibinfo {pages} {141} (\bibinfo
{year} {1990})}\BibitemShut {NoStop}%
\bibitem [{\citenamefont {Avron}\ and\ \citenamefont
	{Kons}(1999)}]{Avron_JPA99}%
\BibitemOpen
\bibfield  {author} {\bibinfo {author} {\bibfnamefont {J.~E.}\ \bibnamefont
		{Avron}}\ and\ \bibinfo {author} {\bibfnamefont {Z.}~\bibnamefont {Kons}},\
}\href@noop {} {\bibfield  {journal} {\bibinfo  {journal} {J. Phys. A-Math.
		Gen.}\ }\textbf {\bibinfo {volume} {32}},\ \bibinfo {pages} {6097} (\bibinfo
{year} {1999})}\BibitemShut {NoStop}%
\bibitem [{\citenamefont {Rice}\ and\ \citenamefont {Mele}(1982)}]{Rice_PRL82}%
\BibitemOpen
\bibfield  {author} {\bibinfo {author} {\bibfnamefont {M.}~\bibnamefont
		{Rice}}\ and\ \bibinfo {author} {\bibfnamefont {E.}~\bibnamefont {Mele}},\
}\href@noop {} {\bibfield  {journal} {\bibinfo  {journal} {Phys. Rev. Lett.}\
}\textbf {\bibinfo {volume} {49}},\ \bibinfo {pages} {1455} (\bibinfo {year}
{1982})}\BibitemShut {NoStop}%
\bibitem [{\citenamefont {Xiao}\ \emph {et~al.}(2010)\citenamefont {Xiao},
	\citenamefont {Chang},\ and\ \citenamefont {Niu}}]{xiao2010berry}%
\BibitemOpen
\bibfield  {author} {\bibinfo {author} {\bibfnamefont {D.}~\bibnamefont
		{Xiao}}, \bibinfo {author} {\bibfnamefont {M.-C.}\ \bibnamefont {Chang}}, \
	and\ \bibinfo {author} {\bibfnamefont {Q.}~\bibnamefont {Niu}},\ }\href@noop
{} {\bibfield  {journal} {\bibinfo  {journal} {Rev. Mod. Phys.}\ }\textbf
	{\bibinfo {volume} {82}},\ \bibinfo {pages} {1959} (\bibinfo {year}
	{2010})}\BibitemShut {NoStop}%
\bibitem [{\citenamefont {Ferrari}(1998)}]{Ferrari_IJMP98}%
\BibitemOpen
\bibfield  {author} {\bibinfo {author} {\bibfnamefont {R.}~\bibnamefont
		{Ferrari}},\ }\href@noop {} {\bibfield  {journal} {\bibinfo  {journal} {Int.
			J. Mod. Phys. B}\ }\textbf {\bibinfo {volume} {12}},\ \bibinfo {pages} {1105}
	(\bibinfo {year} {1998})}\BibitemShut {NoStop}%
\bibitem [{\citenamefont {Kitagawa}\ \emph {et~al.}(2010)\citenamefont
	{Kitagawa}, \citenamefont {Berg}, \citenamefont {Rudner},\ and\ \citenamefont
	{Demler}}]{kitagawa2010topological}%
\BibitemOpen
\bibfield  {author} {\bibinfo {author} {\bibfnamefont {T.}~\bibnamefont
		{Kitagawa}}, \bibinfo {author} {\bibfnamefont {E.}~\bibnamefont {Berg}},
	\bibinfo {author} {\bibfnamefont {M.}~\bibnamefont {Rudner}}, \ and\ \bibinfo
	{author} {\bibfnamefont {E.}~\bibnamefont {Demler}},\ }\href@noop {}
{\bibfield  {journal} {\bibinfo  {journal} {Phys. Rev. B}\ }\textbf {\bibinfo
		{volume} {82}},\ \bibinfo {pages} {235114} (\bibinfo {year}
	{2010})}\BibitemShut {NoStop}%
\bibitem [{\citenamefont {Russomanno}\ \emph {et~al.}(2011)\citenamefont
	{Russomanno}, \citenamefont {Pugnetti}, \citenamefont {Brosco},\ and\
	\citenamefont {Fazio}}]{Russomanno_PRB11}%
\BibitemOpen
\bibfield  {author} {\bibinfo {author} {\bibfnamefont {A.}~\bibnamefont
		{Russomanno}}, \bibinfo {author} {\bibfnamefont {S.}~\bibnamefont
		{Pugnetti}}, \bibinfo {author} {\bibfnamefont {V.}~\bibnamefont {Brosco}}, \
	and\ \bibinfo {author} {\bibfnamefont {R.}~\bibnamefont {Fazio}},\
}\href@noop {} {\bibfield  {journal} {\bibinfo  {journal} {Phys. Rev. B}\
}\textbf {\bibinfo {volume} {83}},\ \bibinfo {pages} {214508} (\bibinfo
{year} {2011})}\BibitemShut {NoStop}%
\bibitem [{\citenamefont {Sambe}(1973)}]{sambe1973steady}%
\BibitemOpen
\bibfield  {author} {\bibinfo {author} {\bibfnamefont {H.}~\bibnamefont
		{Sambe}},\ }\href@noop {} {\bibfield  {journal} {\bibinfo  {journal} {Phys.
			Rev. A}\ }\textbf {\bibinfo {volume} {7}},\ \bibinfo {pages} {2203} (\bibinfo
	{year} {1973})}\BibitemShut {NoStop}%
\bibitem [{\citenamefont {Shirley}(1965)}]{Shirley_PR65}%
\BibitemOpen
\bibfield  {author} {\bibinfo {author} {\bibfnamefont {J.~H.}\ \bibnamefont
		{Shirley}},\ }\href {\doibase 10.1103/PhysRev.138.B979} {\bibfield  {journal}
	{\bibinfo  {journal} {Phys. Rev.}\ }\textbf {\bibinfo {volume} {138}},\
	\bibinfo {pages} {B979} (\bibinfo {year} {1965})}\BibitemShut {NoStop}%
\bibitem [{\citenamefont {Neumann}\ and\ \citenamefont
	{Wigner}(1929)}]{wigner_1929}%
\BibitemOpen
\bibfield  {author} {\bibinfo {author} {\bibfnamefont {J.~V.}\ \bibnamefont
		{Neumann}}\ and\ \bibinfo {author} {\bibfnamefont {E.}~\bibnamefont
		{Wigner}},\ }\href@noop {} {\bibfield  {journal} {\bibinfo  {journal} {Z.
			Phys.}\ }\textbf {\bibinfo {volume} {30}},\ \bibinfo {pages} {467} (\bibinfo
	{year} {1929})}\BibitemShut {NoStop}%
\bibitem [{\citenamefont {Lindner}\ \emph {et~al.}(2017)\citenamefont
	{Lindner}, \citenamefont {Berg},\ and\ \citenamefont
	{Rudner}}]{Lindner_PRX17}%
\BibitemOpen
\bibfield  {author} {\bibinfo {author} {\bibfnamefont {N.~H.}\ \bibnamefont
		{Lindner}}, \bibinfo {author} {\bibfnamefont {E.}~\bibnamefont {Berg}}, \
	and\ \bibinfo {author} {\bibfnamefont {M.~S.}\ \bibnamefont {Rudner}},\
}\href@noop {} {\bibfield  {journal} {\bibinfo  {journal} {Phys. Rev. X}\
}\textbf {\bibinfo {volume} {7}},\ \bibinfo {pages} {011018} (\bibinfo {year}
{2017})}\BibitemShut {NoStop}%
\bibitem [{\citenamefont {Russomanno}\ \emph {et~al.}(2012)\citenamefont
	{Russomanno}, \citenamefont {Silva},\ and\ \citenamefont
	{Santoro}}]{Russomanno_PRL12}%
\BibitemOpen
\bibfield  {author} {\bibinfo {author} {\bibfnamefont {A.}~\bibnamefont
		{Russomanno}}, \bibinfo {author} {\bibfnamefont {A.}~\bibnamefont {Silva}}, \
	and\ \bibinfo {author} {\bibfnamefont {G.~E.}\ \bibnamefont {Santoro}},\
}\href@noop {} {\bibfield  {journal} {\bibinfo  {journal} {Phys. Rev. Lett.}\
}\textbf {\bibinfo {volume} {109}},\ \bibinfo {pages} {257201} (\bibinfo
{year} {2012})}\BibitemShut {NoStop}%
\bibitem [{\citenamefont {Lazarides}\ \emph {et~al.}(2014)\citenamefont
	{Lazarides}, \citenamefont {Das},\ and\ \citenamefont
	{Moessner}}]{Lazarides_PRL14}%
\BibitemOpen
\bibfield  {author} {\bibinfo {author} {\bibfnamefont {A.}~\bibnamefont
		{Lazarides}}, \bibinfo {author} {\bibfnamefont {A.}~\bibnamefont {Das}}, \
	and\ \bibinfo {author} {\bibfnamefont {R.}~\bibnamefont {Moessner}},\
}\href@noop {} {\bibfield  {journal} {\bibinfo  {journal} {Phys. Rev. Lett.}\
}\textbf {\bibinfo {volume} {112}},\ \bibinfo {pages} {150401} (\bibinfo
{year} {2014})}\BibitemShut {NoStop}%
\bibitem [{Note_Cohen({\natexlab{b}})}]{Note_Supp}%
\BibitemOpen
\bibinfo {note} {See Supplementary Material for a detailed derivation of the
	Floquet diagonal ensemble value of the pumped charge, which includes
	Ref.~\cite{Avron_CMP87}.}\BibitemShut {Stop}%
\bibitem [{\citenamefont {Avron}\ \emph {et~al.}(1987)\citenamefont {Avron},
	\citenamefont {Seiler},\ and\ \citenamefont {Yaffe}}]{Avron_CMP87}%
\BibitemOpen
\bibfield  {author} {\bibinfo {author} {\bibfnamefont {J.}~\bibnamefont
		{Avron}}, \bibinfo {author} {\bibfnamefont {R.}~\bibnamefont {Seiler}}, \
	and\ \bibinfo {author} {\bibfnamefont {L.}~\bibnamefont {Yaffe}},\
}\href@noop {} {\bibfield  {journal} {\bibinfo  {journal} {Communications in
		Mathematical Physics}\ }\textbf {\bibinfo {volume} {110}},\ \bibinfo {pages}
{33} (\bibinfo {year} {1987})}\BibitemShut {NoStop}%
\bibitem [{\citenamefont {Rigolin}\ \emph {et~al.}(2008)\citenamefont
	{Rigolin}, \citenamefont {Ortiz},\ and\ \citenamefont
	{Ponce}}]{Rigolin_PRA08}%
\BibitemOpen
\bibfield  {author} {\bibinfo {author} {\bibfnamefont {G.}~\bibnamefont
		{Rigolin}}, \bibinfo {author} {\bibfnamefont {G.}~\bibnamefont {Ortiz}}, \
	and\ \bibinfo {author} {\bibfnamefont {V.~H.}\ \bibnamefont {Ponce}},\
}\href@noop {} {\bibfield  {journal} {\bibinfo  {journal} {Phys. Rev. A}\
}\textbf {\bibinfo {volume} {78}},\ \bibinfo {pages} {052508} (\bibinfo
{year} {2008})}\BibitemShut {NoStop}%
\bibitem [{\citenamefont {Young}\ and\ \citenamefont
	{Deal~Jr}(1970)}]{Young_JMP70}%
\BibitemOpen
\bibfield  {author} {\bibinfo {author} {\bibfnamefont {R.~H.}\ \bibnamefont
		{Young}}\ and\ \bibinfo {author} {\bibfnamefont {W.~J.}\ \bibnamefont
		{Deal~Jr}},\ }\href@noop {} {\bibfield  {journal} {\bibinfo  {journal} {J.
			Math. Phys.}\ }\textbf {\bibinfo {volume} {11}},\ \bibinfo {pages} {3298}
	(\bibinfo {year} {1970})}\BibitemShut {NoStop}%
\bibitem [{\citenamefont {Breuer}\ and\ \citenamefont
	{Holthaus}(1989)}]{breuer1989adiabatic}%
\BibitemOpen
\bibfield  {author} {\bibinfo {author} {\bibfnamefont {H.}~\bibnamefont
		{Breuer}}\ and\ \bibinfo {author} {\bibfnamefont {M.}~\bibnamefont
		{Holthaus}},\ }\href@noop {} {\bibfield  {journal} {\bibinfo  {journal} {Z.
			Phys. D}\ }\textbf {\bibinfo {volume} {11}},\ \bibinfo {pages} {1} (\bibinfo
	{year} {1989})}\BibitemShut {NoStop}%
\bibitem [{\citenamefont {Eckardt}\ \emph {et~al.}(2005)\citenamefont
	{Eckardt}, \citenamefont {Weiss},\ and\ \citenamefont
	{Holthaus}}]{Eckardt_PRL05}%
\BibitemOpen
\bibfield  {author} {\bibinfo {author} {\bibfnamefont {A.}~\bibnamefont
		{Eckardt}}, \bibinfo {author} {\bibfnamefont {C.}~\bibnamefont {Weiss}}, \
	and\ \bibinfo {author} {\bibfnamefont {M.}~\bibnamefont {Holthaus}},\
}\href@noop {} {\bibfield  {journal} {\bibinfo  {journal} {Phys. Rev. Lett.}\
}\textbf {\bibinfo {volume} {95}},\ \bibinfo {pages} {260404} (\bibinfo
{year} {2005})}\BibitemShut {NoStop}%
\bibitem [{\citenamefont {Russomanno}\ and\ \citenamefont
	{Dalla~Torre}(2016)}]{Russomanno_EPL16}%
\BibitemOpen
\bibfield  {author} {\bibinfo {author} {\bibfnamefont {A.}~\bibnamefont
		{Russomanno}}\ and\ \bibinfo {author} {\bibfnamefont {E.~G.}\ \bibnamefont
		{Dalla~Torre}},\ }\href@noop {} {\bibfield  {journal} {\bibinfo  {journal}
		{Europhys. Lett.}\ }\textbf {\bibinfo {volume} {115}},\ \bibinfo {pages}
	{30006} (\bibinfo {year} {2016})}\BibitemShut {NoStop}%
\bibitem [{\citenamefont {Wauters}\ and\ \citenamefont
	{Santoro}()}]{Wauters_unpub}%
\BibitemOpen
\bibfield  {author} {\bibinfo {author} {\bibfnamefont {M.~M.}\ \bibnamefont
		{Wauters}}\ and\ \bibinfo {author} {\bibfnamefont {G.~E.}\ \bibnamefont
		{Santoro}},\ }\href@noop {} {}\bibinfo {note} {(in preparation)}\BibitemShut
{NoStop}%
\bibitem [{\citenamefont {Privitera}\ and\ \citenamefont
	{Santoro}(2016)}]{Privitera_PRB16}%
\BibitemOpen
\bibfield  {author} {\bibinfo {author} {\bibfnamefont {L.}~\bibnamefont
		{Privitera}}\ and\ \bibinfo {author} {\bibfnamefont {G.~E.}\ \bibnamefont
		{Santoro}},\ }\href@noop {} {\bibfield  {journal} {\bibinfo  {journal} {Phys.
			Rev. B}\ }\textbf {\bibinfo {volume} {93}},\ \bibinfo {pages} {241406}
	(\bibinfo {year} {2016})}\BibitemShut {NoStop}%
\end{thebibliography}
\end{document}


\title{Supplementary Material for\\Non-adiabatic breaking of topological pumping}

\author{Lorenzo Privitera}
\affiliation{SISSA, Via Bonomea 265, I-34136 Trieste, Italy}
\author{Angelo Russomanno}
\affiliation{Scuola Normale Superiore, Piazza dei Cavalieri 7, I-56127 Pisa, Italy}
\affiliation{International Centre for Theoretical Physics (ICTP), P.O.Box 586, I-34014 Trieste, Italy}

\author{Roberta Citro}
 \affiliation{Dipartimento di Fisica "E.R. Caianiello", Universit\`a
  degli Studi di Salerno and Spin-CNR Unit\`a, Via Giovanni Paolo II, 132, I-84084 Fisciano
  (Sa), Italy}

\author{Giuseppe E. Santoro}
\affiliation{SISSA, Via Bonomea 265, I-34136 Trieste, Italy}
\affiliation{International Centre for Theoretical Physics (ICTP), P.O.Box 586, I-34014 Trieste, Italy}
\affiliation{CNR-IOM Democritos National Simulation Center, Via Bonomea 265, I-34136 Trieste, Italy}

\maketitle

\subsection{Derivation of the long-time pumped charge formula.}
In this section we derive the expression for the charge pumped in a cycle in the long-time limit, Eq.~(4) of the main text. 
We start from the definition of pumped charge over $m$ periods
\begin{equation*}
Q(m\tau) = \frac{1}{L} \int_{0}^{m\tau} \Tr \tonde{\hat{\rho}(t) \hat{J}(t)}  dt \;,
\end{equation*} 
where $ \hat{\rho}(t)$ is the instantaneous density matrix of the system in the Schr\"odinger picture and $\hat{J}(t)=\partial_{\kappa} \hat{H}(t)/\hbar$ is the corresponding current operator. As we said in the main text, we consider a unitary dynamics and a generic initial density matrix $\hat{\rho}(0)$. We stress the fact that we are using a Schr\"{o}dinger representation: the only time dependence of the current operator is the explicit one.  Using the formula $\hat{J}(t)=\partial_{\kappa} \hat{H}(t)/\hbar$ we can write the pumped charge as
\begin{equation}
\begin{split}
Q(m\tau) = \frac{1}{\hbar L}  \int_{0}^{m\tau} \!\mathrm{d}t\; \Tr \biggl(\hat{\rho}(0) \hat{U}^{\dagger}(m\tau,0) \times \\ \times\bigl[\partial_{\kappa} \hat{H}(t)\bigr] \hat{U}(m\tau,0) \biggr)   \;.
\end{split}
\end{equation} 
Making use of the following equality (see Refs.~\cite{Avron_CMP87,kitagawa2010topological})
\begin{equation}\label{eq:DerEvOp}
\partial_{\kappa} \hat{U}(\tau, 0) = -\frac{i}{\hbar} \int_{0}^{\tau} \mathrm{d}t \; \hat{U}(\tau,t) \quadre{ \partial_{\kappa} \Ham(t)} \hat{U}(t,0)\;,
\end{equation}
and of the fundamental property $\hat{U}^\dagger(t,0) = \hat{U}(0,t) = \hat{U}(0,\tau) \hat{U}(\tau,t)$, we arrive at
\begin{equation}\label{eq:charge2}
Q(m\tau) = \frac{1}{L}\Tr \tonde{\hat{\rho}(0) \; \hat{U}^{-1}(m\tau,0)  \left[i \partial_{\kappa} \hat{U}(m\tau,0)\right]} \; .
\end{equation}
Now we can use Floquet theorem, namely 
\begin{equation}
\begin{split}
\hat{U}(m\tau,0) =& \left[\hat{U}(\tau,0)\right]^m =\\ \nonumber= &\sum_{\alpha}\nep^{-\frac{i}{\hbar} m\varepsilon_{\alpha} \tau}\ket{\phi_{\alpha}(0)}\bra{\phi_{\alpha}(0)} \;.
\end{split}
\end{equation}
Substituting in Eq.~\eqref{eq:charge2}, we obtain
\begin{equation}\label{eq:ChargeLongTime}
\begin{split}
Q(m\tau)= & \frac{1}{L}\Tr 
\biggl\lbrace\hat{\rho}(0)  \sum_{\alpha} \biggl( \frac{m\tau}{\hbar} \partial_\kappa  \varepsilon_{\alpha} \ket{\phi_{\alpha}(0)}\bra{\phi_{\alpha}(0)}
+ \\ 
 +& \sum_{\alpha' \neq \alpha} \nep^{\frac{i}{\hbar} \left(\varepsilon_{\alpha'} - \varepsilon_{\alpha} \right)m\tau} \times \\
 \times & \ket{\phi_{\alpha'}(0)} \bra{\phi_{\alpha'}(0)} \partial_\kappa \phi_{\alpha}(0) \rangle \bra{\phi_{\alpha}(0)}\biggr) \biggr\rbrace .
\end{split}
\end{equation}
We now specialize this expression to the case where the system is translationally invariant and non-interacting. 
The density matrix factorizes as $\hat{\rho} = \prod_{k} \hat{\rho}_k$. In the same way, the Floquet modes and quasienergies will be labelled by single-particle quasimomentum $k$. 
It is then convenient to take the trace in the Floquet basis $\ket{\phi_{\alpha,k}(0)}$, so that
\begin{equation}\label{eq:ChargeLongTime2}
\begin{split}
Q(m\tau)= & \frac{m}{\hbar\omega}  \int_{-\frac{\pi}{a}}^{+\frac{\pi}{a}} dk \;\biggl(\sum_{\alpha} \rho_{\alpha\alpha,k}(0)
\frac{\partial \varepsilon_{\alpha,k}}{\partial k}
+ \\ 
+& \sum_{\alpha' \neq \alpha} \frac{\nep^{\frac{i}{\hbar} \left(\varepsilon_{\alpha'} - \varepsilon_{\alpha} \right)m\tau}}{m} \times \\ \times & \rho_{\alpha\alpha',k}(0)\bra{\phi_{\alpha',k}(0)} 
\partial_\kappa \phi_{\alpha,k}(0) \rangle \biggr) \;.
\end{split}
\end{equation}
We have used the fact that $\varepsilon_{\alpha,k}$ depends on $k+\kappa$.
The diagonal matrix elements $\rho_{\alpha\alpha,k}(0)$ are the occupation number of the $\alpha$-th Floquet mode at the initial time. In the main text, we called them $n^T_{\alpha,k}$ in the case of an initial thermal density matrix at temperature $T$ and $n_{\alpha,k}$ when $T=0$. In a second quantization formalism they can be written as $n^T_{\alpha,k} = \Tr \left( {\hat{\rho}}_{\mathsmaller{T}}\opfdag{\alpha,k} \opf{\alpha,k}\right)$. In the long time limit, e.g. $m\to \infty$, the highly-oscillatory off-diagonal term disappears because of the Riemann-Lebesgue lemma \cite{Russomanno_PRL12}. As a consequence, the average pumped charge converges to the so called \textit{diagonal ensemble} value, that is 
\begin{equation}
\lim_{m\to\infty} \frac{Q(m\tau)}{m} \equiv Q_{\rm diag}   =
\frac{1}{\hbar \omega} \sum_{\alpha} \int_{-\frac{\pi}{a}}^{+\frac{\pi}{a}} \! \mathrm{d}k \; n^T_{\alpha,k}
\frac{\partial \varepsilon_{\alpha,k}}{\partial k}  .
\end{equation}

\bibliography{ThesisBiblio}